# Finite temperature dynamical polarization and plasmons in gapped graphene

**Digish K Patel**[*,1], **Syed S Z Ashraf**[**,2], and **Ami C Sharma**[*,1]

[1] Physics Department, Faculty of Science, The M.S. University of Baroda, Vadodara-390002, Gujarat, India
[2] Physics Department, Faculty of Science, Aligarh Muslim University, Aligarh-202002, Uttar Pradesh, India

\* e-mail jdiggish@gmail.com
\*\* Corresponding author: e-mail ssz_ashraf@rediffmail.com
\* e-mail acs_phy@yahoo.com

In this study we report our numerical results on finite temperature non-interacting dynamical polarization function, plasmon modes and electron energy loss function of doped single layer gapped graphene within the random phase approximation. We find that the interplay of linear energy band dispersion, chirality, bandgap and temperature endow single layer gapped graphene with strange polarizability behaviour which is a mixture of 2DEG, single layer and bilayer graphene and as a result the plasmon spectrum also manifests strikingly peculiar behaviour. The plasmon dispersion is observed to be suppressed till temperatures up to ~ $0.5T_f$, similar to the gapless graphene case but beyond $0.5T_f$ a reversal in trend is seen in gapped graphene, for all values of band gap. This behaviour is also corroborated by the density plots of electron energy loss function. The opening of a small gap also generates a new undamped plasmon mode which is found to disappear at high temperatures. The plasmonic behaviour of gapped graphene is further found to be hugely influenced by the substrate on which the gapped graphene sheet rests, which signifies the need for a careful substrate selection in the making of desirable graphene based plasmonics devices.

**1 Introduction** Ever since its discovery, Graphene continues to evoke immense interest both theoretically and experimentally due to its amazing structural, mechanical, electrical, electronic, thermal, magnetic and optical properties, warranting a plethora of exciting applications [1]. Amid the multitude of diverse applications of Graphene for novel technologies, one highly promising and rapidly emerging area is of plasmonics where this unique 2D electronically and optically tunable material has cutting edge over the conventional metal plasmonic materials, that makes graphene the choicest material for applications in photonic and optoelectronic technologies [2]. This stems largely from the extremely high atomistic confinement of electrons and the small spatial spread of the associated electromagnetic field; of the order of a million times smaller than the wavelength of light, enabling improved imaging resolution, strong light-matter interaction, a relatively low loss in THz and infrared region frequencies, wide tunability of plasmon frequencies through electrical or chemical modification of the charge density, ranging from terahertz to mid-infrared and long lived plasmon lifetimes -overcoming the limitations of noble metal plasmonics [3]. The envisaged applications of graphene plasmonics are overwhelmingly impressive, covering diverse areas such as electronics, linear and non-linear optics, spectroscopy, THz technology, energy storage, biotechnology, cancer therapy, and many others [3-6]. Already there have been quite a number of reviews [2-12] in a short span of this fast advancing sub-field of Graphene plasmonics which speaks about the vigorous research that is being pursued in this field and in no matter of time the experimental and theoretical proliferation of research papers calls for a fresh topical review.

Surface Plasmons in gapless graphene are best suited to Tera-Hertz (THz) and far infrared frequencies while metals plasmonics excel in the visible and near infrared region [2]. This restriction in the operating plasmonic frequency of graphene confined to THz and far infrared frequencies is a setback in the propagation of Plasmons as still the deprivation of reliable THz sources and detectors hinder graphene plasmonics applications [4]. This inadequacy coupled with smaller propagation length to wavelength ratio at high frequencies of the surface plasmons limits graphene from outsmarting conventional noble metal based plasmonics. Some of the speculated ways of extending the operating graphene plasmonic frequencies to visible and near infrared and thereby increasing the scope of prospective applications are by excessive non-destructive doping, metal hybrid structures and the opening of band gap [4]. Recently, some experiments based on angle-resolved photoemission spectroscopy (ARPES) have shown the opening of band gap of $\approx 260\ meV\ \&\ \approx 53\ meV$ on graphene sample epitaxially grown on SiC (Silicon Carbide) & BN (Boron Nitride) substrates, respectively. This gap opening is due to the symmetry breaking of A and B sublattices arising due to interaction between substrate and graphene sheet [13-16]. Also a small gap ($\sim 10^{-3}$ meV) in the absence of any substrate but arising to the spin-orbit interaction has also been observed [17].

Collective excitations of graphene based systems like Single Layer Graphene (SLG) [18-19], Bilayer Graphene (BLG) [20-24], Single Layer Gapped Graphene (SLGG) [25] have been studied extensively in the recent past. For studying the collective excitations, screening and other many body effects the central quantity of interest is the polarization function. The non-interacting dynamical polarization function (NDP) and plasmon dispersions have been extensively studied in past for undoped and doped gapless graphene at zero temperature [18-19] as well as at finite temperature [26-28] and in a magnetic field [29-30]. The undoped graphene cannot sustain low energy plasmon modes at zero temperature but at finite temperature plasmons exist because of finite electron density arising due to thermal broadening of Fermi function [27] while in doped graphene plasmon dispersion is supported at zero-temperature as well.

All the reported calculated analytical and numerical expressions of doped SLGG are restricted to the case of zero-temperature [25, 31-32] whereas the experimental study of plasmon dispersion is carried out at a finite temperature and finite collision rate. Though, one study has reported the polarization function and plasmons for SLGG but at very small value of the energy gap (0.08 meV) and at very low temperatures (2K) [33]. This small gap is generated by the spin orbit interaction in graphene and without breaking the symmetry of the lattice. Another difference is that the model Hamiltonian used in that study contains an additional gap dependent term due to spin orbital interaction which respects all of the symmetries of graphene. In this paper, we assume a gap produced by the symmetry breaking of A and B sublattices, arising due to interaction between substrate and graphene sheet, and a gap dependent Hamiltonian breaking the mirror symmetry. The temperature effects on NDP function and plasmon dispersion in doped SLGG are still unclear and need a systematic investigation. In particular, the thermal effects are expected to be significant for larger $k_B T/\varepsilon_f$ ratio and smaller dielectric constant value of the substrate, where $\varepsilon_f = \hbar v_f k_f$ is the Fermi energy of SLG with $v_f \approx 10^6$m/s is Fermi velocity and $k_f = (\pi n)^{1/2}$ is

Fermi momentum ($n$ is carrier concentration) [26]. Experimentally, the Plasmons in metals and graphene are determined through high resolution Electron Energy Loss Spectroscopy (EELS) which yields the electron energy loss function. This quantity is proportional to the imaginary part of inverse dielectric function of SLGG. We therefore in this paper investigate the effects of finite temperature on the behaviour of plasmons in SLGG.

The paper is organized thus: Section 2 briefly discusses the formalism used for the calculating the gapped graphene temperature dependent polarization function; in section 3 we report the results and discussion on (3.1) polarization function, (3.2) plasmon dispersion and (3.3) energy loss spectrum of SLGG and finally, the study is concluded in section 4.

## 2 Formalism

The temperature dependent dynamic dielectric function for single layer gapped graphene (SLGG) within RPA can be written as [18],

$$\epsilon_{\text{SLGG}}(q, \omega, T) = 1 - \frac{2\pi e^2}{\kappa q} \Pi_{\text{SLGG}}(q, \omega, T) \qquad (1)$$

where $e$ is the electronic charge, $q$ is the wave vector, $\kappa = 5.5$ is the average background dielectric constant of graphene placed on SiC with the other side being exposed to air [31], $\Pi_{\text{SLGG}}(q, \omega, T)$ is the finite temperature dynamic polarizability given by [18, 34],

$$\Pi_{\text{SLGG}}(q, \omega, T) = \frac{g}{L^2} \sum_{k,s,s'} \frac{(f_{sk} - f_{s'k'})}{\hbar\omega + E_{sk} - E_{s'k'} + i\,\hbar/\tau} F_{ss'}, \quad (2)$$

where g is the degeneracy factor, $L$ is the area of the system, $E_{sk}\left(E_{s'k'}\right) = s(s')\sqrt{[\gamma\,k(k')]^2 + \Delta^2}$ is the energy in the low energy Dirac model, with $\boldsymbol{k}\,(\boldsymbol{k}' = \boldsymbol{k} + \boldsymbol{q})$ being the momentum, $s(s') = \pm 1$ denotes the band index corresponding to the conduction band ($+$) and valance band ($-$), respectively, $\gamma = 6.46$ eV Å is band parameter, $2\Delta$ is the bandgap, $F_{ss'}$ is the characteristic wave-function overlap factor define as [31],

$$F_{ss'} = \frac{1}{2}\left[1 + \frac{(ss')}{\sqrt{(\gamma\,k)^2 + \Delta^2}}\left(\sqrt{(\gamma\,k)^2 + \Delta^2} + \frac{\gamma^2\,k\,q\,\cos\phi}{\sqrt{(\gamma\,k)^2 + \Delta^2}}\right)\right],$$

$$(3)$$

where $\phi$ is the angle between $\boldsymbol{k}$ and $\boldsymbol{q}$, $\tau$ is relaxation time- which is a very important parameter because its value affects the plasmon propagation distance. We take $\tau$ corresponding to DC mobility $\mu \approx 10000$ cm$^2$/(V s) and $\varepsilon_f \approx 0.8$ eV is $\tau = \mu\varepsilon_f/ev_f^2 \approx 0.8$ picosecond (ps) [5]. The $f_{sk}$ is Fermi distribution function;

$$f_{sk} = \left[1 + e^{[(E_{sk} - \mu_c)\beta]}\right]^{-1}.\quad(4)$$

Here $\mu_c$ is the finite temperature chemical potential determined by the conservation of the total electron density, and defined through [35],

$$n = \int_\Delta^\infty dE_{sk}\left\{\frac{N(E_{sk})}{[1 + e^{[(E_{sk} - \mu_c)\beta]}]} - \frac{N(E_{sk})}{[1 + e^{[(E_{sk} + \mu_c)\beta]}]}\right\}.\quad(5)$$

Where $\beta = 1/k_b T$, $N(E_{sk}) = 2E_{sk}/(\pi\gamma^2)$ is the density of states of SLGG and $n = \varepsilon_f^2/(\pi\gamma^2)$ is the net carrier concentration. The eq. (2) can be rewritten to get the real and imaginary part of dynamic polarizability at a finite temperature as,

$$\text{Im}[\Pi_{\text{SLGG}}(q, \omega, T)] = -\frac{g}{2}\sum_{ss'}\left\{\int_0^\infty \frac{d^2k}{(2\pi)^2}\right.$$
$$\left.\times \frac{(f_{sk} - f_{s'|k+q|})(\hbar/\tau)(F_{ss'})}{\left[\hbar\omega + s\sqrt{(\gamma k)^2 + \Delta^2} - s'\sqrt{(\gamma|k+q|)^2 + \Delta^2}\right]^2 + (\hbar/\tau)^2}\right\}$$
(6)

and

$$\text{Re}[\Pi_{\text{SLGG}}(q, \omega, T)] = \frac{g}{2}\sum_{ss'}\left\{\int_0^\infty \frac{d^2k}{(2\pi)^2}(f_{sk} - f_{s'|k+q|})\right.$$
$$\left.\times \frac{\left[\hbar\omega + s\sqrt{(\gamma k)^2 + \Delta^2} - s'\sqrt{(\gamma|k+q|)^2 + \Delta^2}\right](F_{ss'})}{\left[\hbar\omega + s\sqrt{(\gamma k)^2 + \Delta^2} - s'\sqrt{(\gamma|k+q|)^2 + \Delta^2}\right]^2 + (\hbar/\tau)^2}\right\}.$$
(7)

## 3 Numerical results and discussion

### 3.1 Polarization function
The numerical solutions of Eq. (6) and Eq. (7) for imaginary and real part of NDP at finite temperature are plotted as a function of frequency in Fig. 1 and Fig. 2, respectively, for a set of values of the scaled band gap parameter $a = \Delta/\mu_f (= 0.3, 0.6 \& 0.9)$ and temperature values ($T = 0, 0.5\,T_f \& T_f$), where $\mu_f = \sqrt{\varepsilon_f^2 + \Delta^2}$ is the Fermi energy of SLGG and $T_f = \varepsilon_f/k_b$ is the Fermi temperature. For zero temperature our computed results (using $\tau = 0.8$ ps) for NDP are in excellent agreement with that reported earlier (where $\tau \to \infty$)[36]. The relaxation time ($\tau$) is an important parameter because the actual value of $\tau$ affects the plasmon propagation distance. For $\tau \to \infty$, the imaginary part of NDP's peak exhibits abrupt step-like behaviour, near the ends of the interval where $\text{Im}[\Pi_{\text{SLGG}}(q, \omega)] = 0$. But for $\tau(= 0.25$ ps & $0.1$ ps), the dips become smoother where $\text{Im}[\Pi_{\text{SLGG}}(q, \omega)] \neq 0$ and the height of the peak also decreases with the decrease in $\tau$, from $0.28$ ps to $0.1$ ps as shown in Fig. 1(g). Here, we extend the calculations further for finite temperatures.

We find that the magnitudes as well as positions of peaks and dips of real and imaginary parts of NDP function is affected considerably at different temperature values. As can be seen from Figs. 1(a)-1(d) and Figs. 2(a)-2(d), the curves of both real and imaginary parts of NDP, for gap values of $(a = 0.3 \& 0.6)$ show comparatively larger peaks and dips at $T = T_f$ and relatively smaller peaks and dips at $T = 0.5T_f$, as compared to that at $T = 0$ K. The peaks and dips at $T = 0$ K are observed to lie in between that at $T = T_f$ and $T = 0.5T_f$. However, for high gap value of $a = 0.9$, the highest and lowest peaks and dips are observed at $T = 0$, as is evident from Figs. 1(e)-1(f) and Figs. 2(e)-2(f), respectively.

In Fig.1, the peaks correspond to single particle excitation region where $\text{Im}[\Pi_{\text{SLGG}}(q, \omega, T)]$ is not equal to zero. At $q = 2.5\,k_f$ and $T = 0$ K, the gap opens between intraband and interband SPE edge and also the gap increases with increasing $a$ (at $a = 0$, intraband SPE edge is equal to interband SPE edge). At finite temperatures, intraband SPE edge shifts toward the higher frequency range and the gap between intraband and interband SPE edge decreases with increasing temperatures because of the fact that wider class of single particle transitions is allowed at higher temperatures.

Also it is noticed from Figs. 1(a), 1(c) & 1(e) that the imaginary part of NDP exhibits single peak for all temperatures in case when $q = 0.5k_f$. However this is true only within the range of $\omega\hbar/\varepsilon_f$ (0 to 1) plotted in figures where only one peak exists. But for $\omega\hbar/\varepsilon_f > 2$, at $T = 0$ K, this value decreases with increasing $T$), interband transitions occur as clearly shown in Fig. 6(a) −upper shaded area. But when $q = 2.5k_f$, it can be observed that two peaks emerge out for all values of temperatures. Moreover, the separation between these two peaks widen with increasing gap values, getting the largest for $a = 0.9$, as noticeable from Fig. 1(f).

In the case of real part of NDP for $q = 0.5k_f$, similar to the imaginary part of NDP single dips are observed for all gap values and temperatures. In accordance with Kramers–Krönig theory, positions and magnitudes of these dips (Figs. 2(a), 2(c) & 2(e)) are closely related with positions and magnitudes of peaks in imaginary part of NDP (Figs. 1(a), 1(c) & 1(e)). In case of the real part of NDP for $q = 2.5k_f$, we get one dip and one peak (Figs. 2(b), 2(d) & 2(f)) related to the sharp cut offs in the imaginary part of NDP (Figs. 1(b), 1(d) & 1(f)) at intraband and interband SPE's boundaries, respectively.

The sign of real part of NDP changes from negative to positive as it sweeps across the SPE region. The positions and magnitudes of these extremums are still related with that of imaginary part of NDP. In Figs. 2(a)-2(f), changes in the sign of real part of NDP from negative to positive indicates a sweep across the electron-hole continuum.

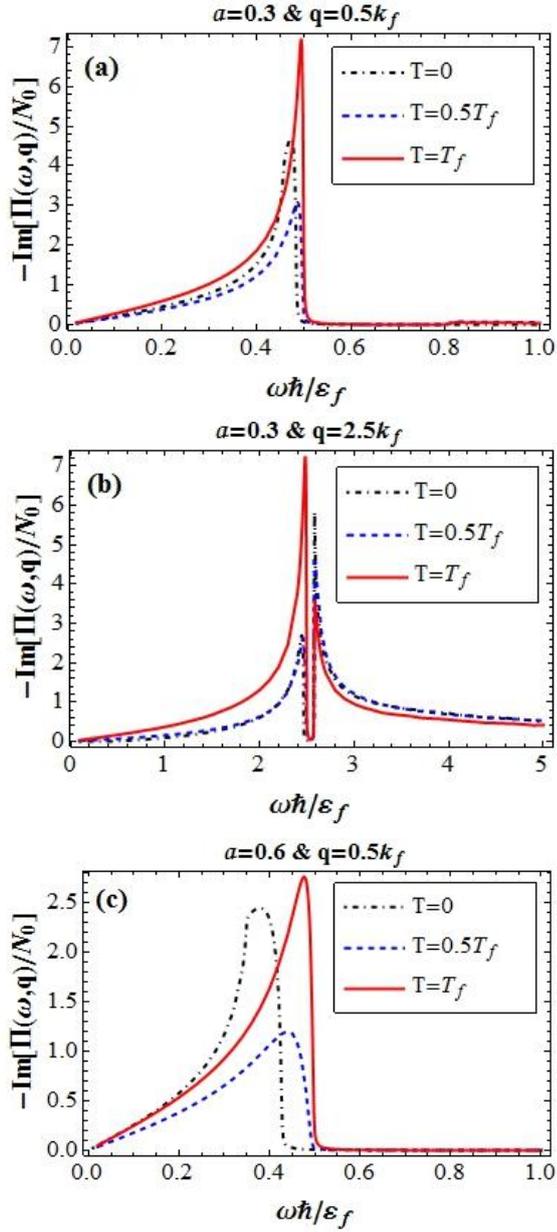

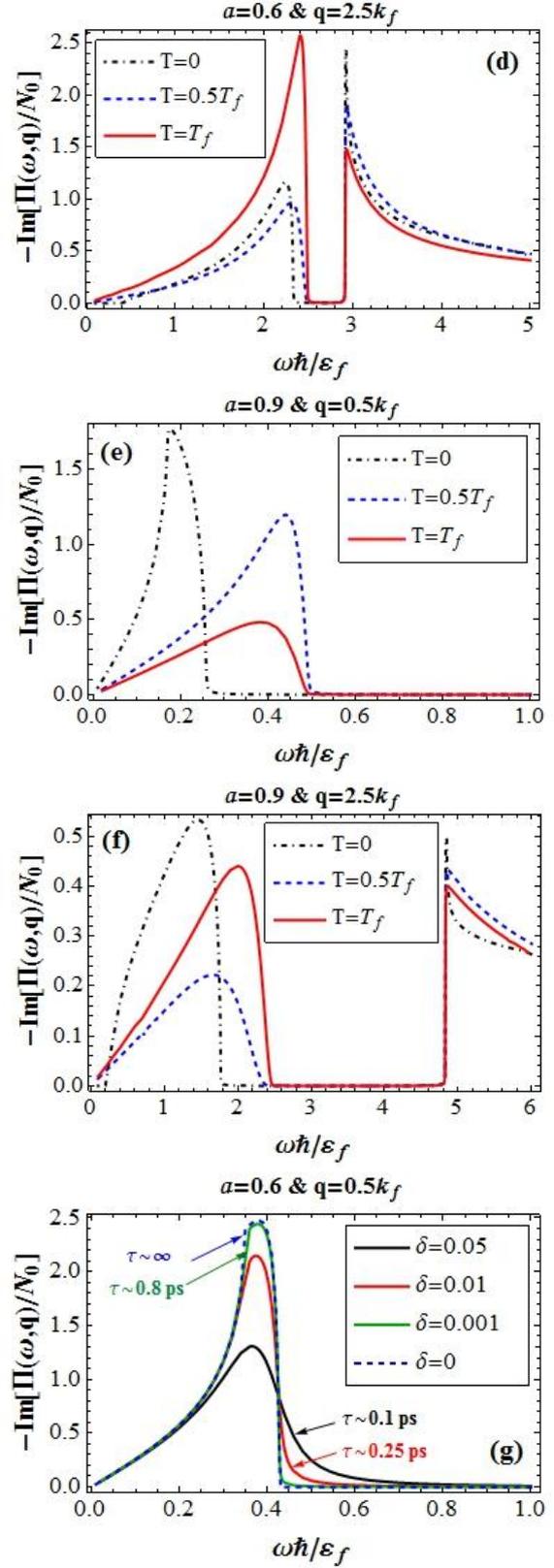

**Figure 1** Temperature dependent imaginary part of the non-interacting polarization function as a function of $\omega$ at different values of wave vectors $q$ using $\delta = 0.001(\tau = 0.8$ ps). For (a), (c), (e) $q = 0.5\,k_F$ and for (b), (d), (f) $q = 2.5\,k_F$. (g) Imaginary part of polarization function as a function of $\omega$ at different values of relaxation time $\tau$ at $T = 0$ K, $a = 0.6$ and $q = 0.5\,k_F$. Here $N_0 = 2\mu_f/(\pi\gamma^2)$ is the density of states at the Fermi level of SLGG [36].

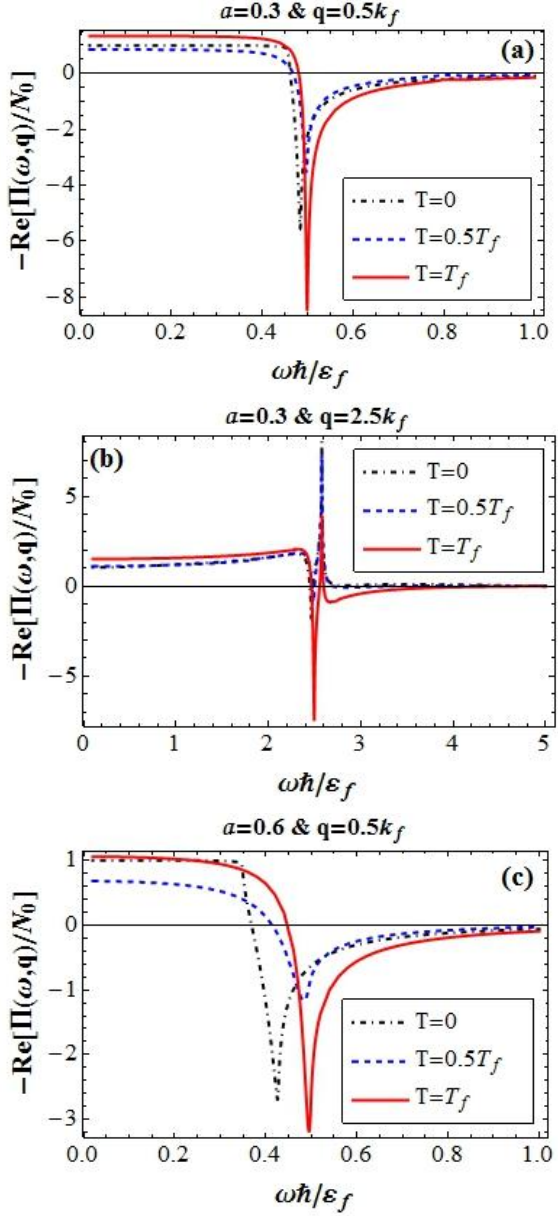

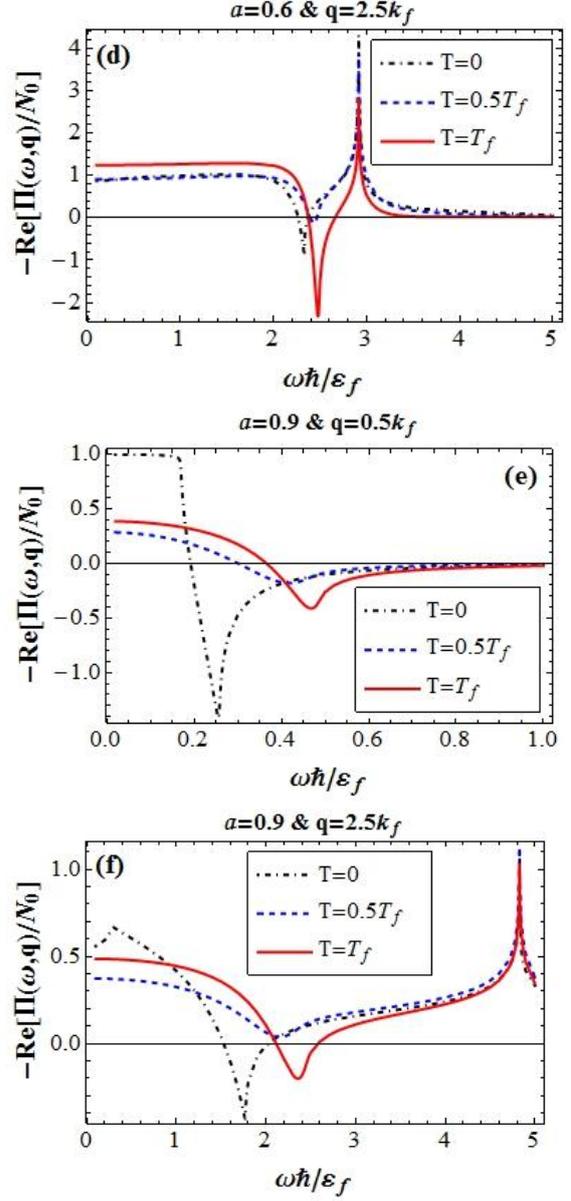

**Figure 2** Temperature dependent real part of the non-interacting polarization function as a function of $\omega$ at wavevector, $q = 0.5\,k_F$ in Figs. (a), (c), (e) and at $q = 2.5\,k_F$ in Figs. (b), (d), (f).

### 3.2 Plasmon dispersion

The plasmon dispersion relation of an electronic system can be defined by the poles of density-density response function, $\Pi_{SLGG}(q, \omega, T)$ or equivalently, from the zeroes of dynamical dielectric function, $\epsilon_{SLGG}(q, \omega, T)$ [25]:

$$\epsilon_{SLGG}(q, \omega_p - i\eta, T) = 0, \tag{8}$$

where, $\omega_p$ is the plasmon frequency at a given wave vector $q$ and $\eta$ is the damping rate of plasma oscillations. In $\omega$

complex plane, if the poles of $\Pi_{SLGG}(q, \omega, T)$ are on the real axis then the plasmons are long-lived and well-defined. However if the poles are away from real axis, we have the Landau damped plasmons due to the electron scattering. In the case of weak damping ($\eta \ll \omega_p$), which is more closer to the real situation, the imaginary part is negligibly small and thus, the plasmon dispersion is given by:

$$\text{Re}[\epsilon_{SLGG}(q, \omega_p, T)] = 0. \tag{9}$$

The numerical solution of Eq. (9) is plotted as a relation between plasmon frequency $\omega_p$ and wave vector $q$ in Figs. 3(a)-3(f) for various values of gap parameter ($a = 0.3, 0.6$ & $0.9$), respectively. We also show the density plot of SPE region where the Landau damping of plasmons takes place calculated from Eq. (6) using $\delta = 0.001$. From Figs. 3(a)-3(f), it is clear that in the long wavelength limit the Plasmon dispersion $\omega_p \propto \sqrt{q}$ behaviour of the 2DEG and gapless graphene is reproduced for SLGG also, but the dispersion for larger values of $q$ exhibits linear behaviour. Also we observe that for all gap values, plasmon frequency increases rapidly with $q$ for higher temperature values as compared to low-temperature values. The rate of increase of $\omega_p$ with respect to wave vector $q$ is observed to be the lowest for $T = 0.5T_f$ for small gap valus($a = 0.3$) but with increasing gap value($a = 0.6$), this rate of increase of $\omega_p$ at $T = 0.5T_f$ case overtakes the curve for $T = 0$ K at around $q \approx 0.6k_f$. However, for highest plotted value of gap($a = 0.9$) at $T = 0.5T_f$, plasmons disappear for $q \gtrsim 0.4k_f$ in the plotted range of $q$ values. For low-temperature, our results on plasmon dispersion curve matches with the experimental values reported earlier [31]. Fig. 3(g) shows an extra undamped plasmon mode that emerges in the gap between interband and intraband SPE regions for $T = 0$ K, $\delta = 10^{-4}$ and $a = 0.05$. This extra undamped plasmon mode disappears at $T = T_f$ and $\delta > 10^{-4}$ because of the enhanced intraband contribution that shifts the peak to below zero. In Fig. 3(h), we display plasmon dispersion relation for various values of coupling constant (defined as $\alpha = 4e^2/\gamma\kappa$ [34]) at $T = 0$ K, $\delta = 10^{-4}$ and $a = 0.05$. We also observe in this figure overlapping extra undamped plasmon modes of Fig. 3(g) at all values of coupling constant for $q \geq k_f$. The reason for overlapping of these extra undamped plasmon modes is the occurrence of extra sharp peaks at the same frequency value for the corresponding wave-vectors for all values of coupling constant, as shown in Fig. 3(j).

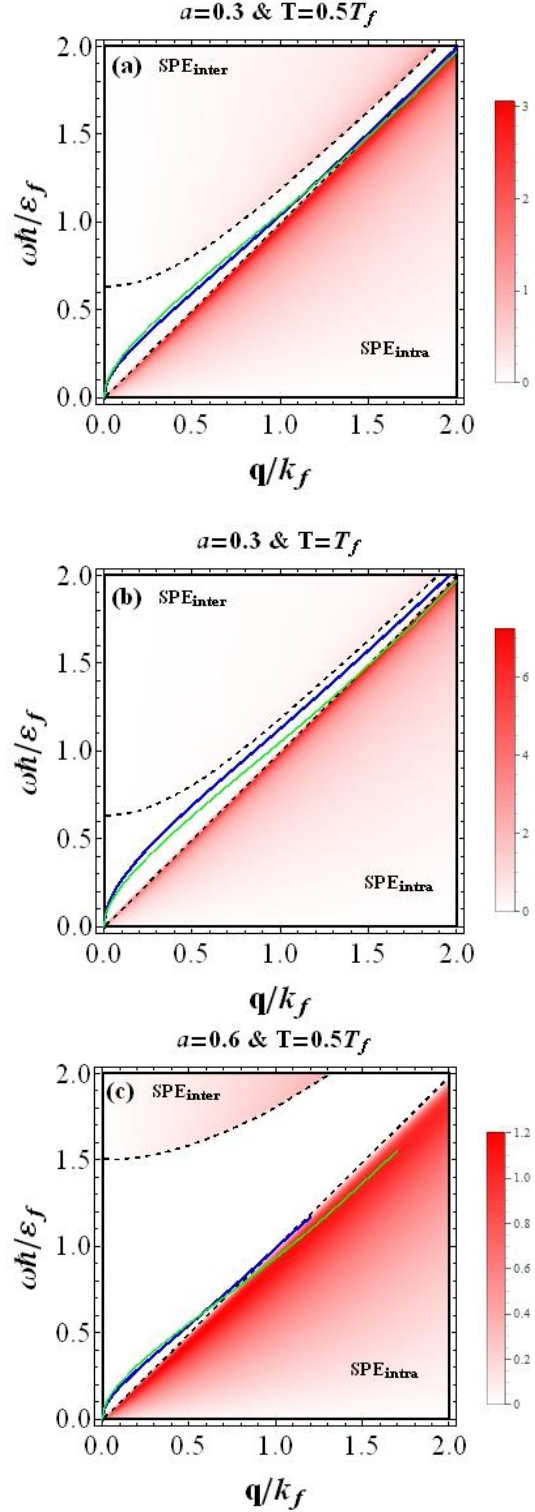

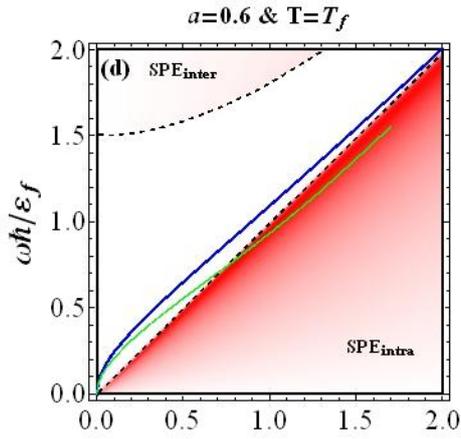

**(d)** $a=0.6$ & $T=T_f$

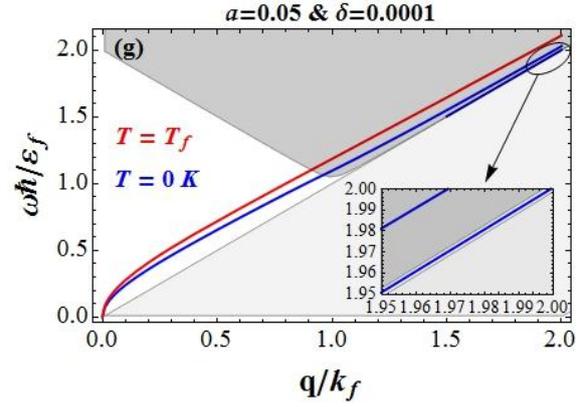

**(g)** $a=0.05$ & $\delta=0.0001$

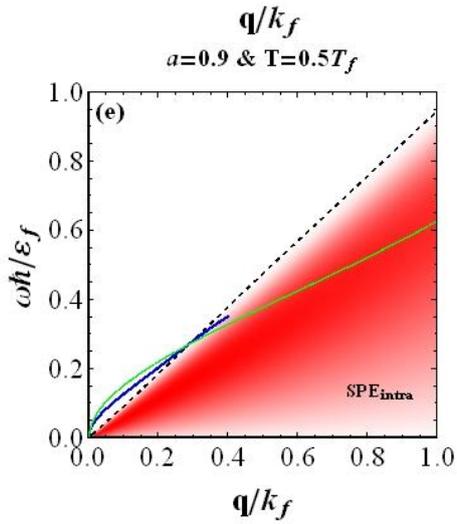

**(e)** $a=0.9$ & $T=0.5T_f$

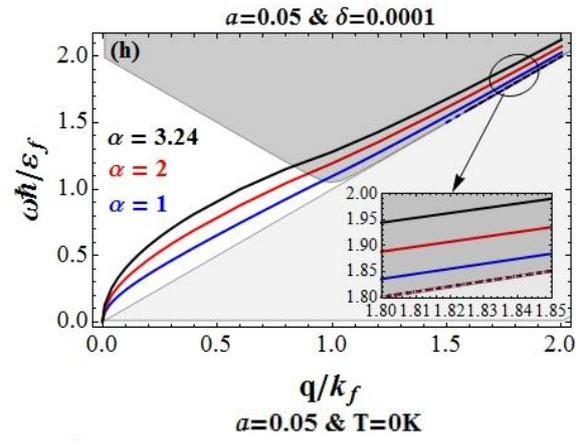

**(h)** $a=0.05$ & $\delta=0.0001$

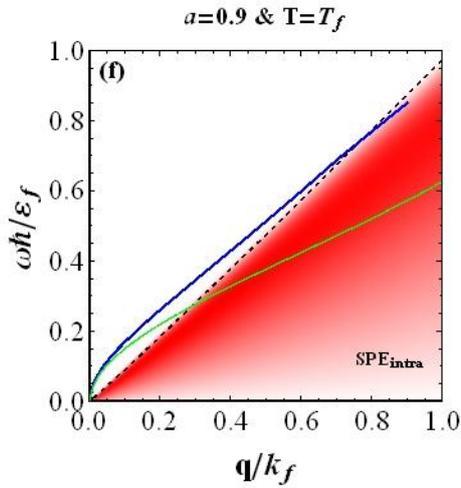

**(f)** $a=0.9$ & $T=T_f$

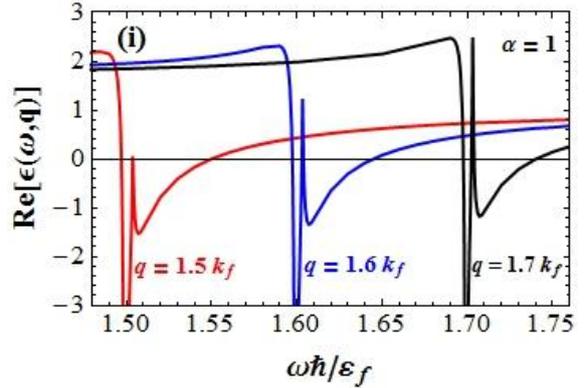

**(i)** $a=0.05$ & $T=0K$

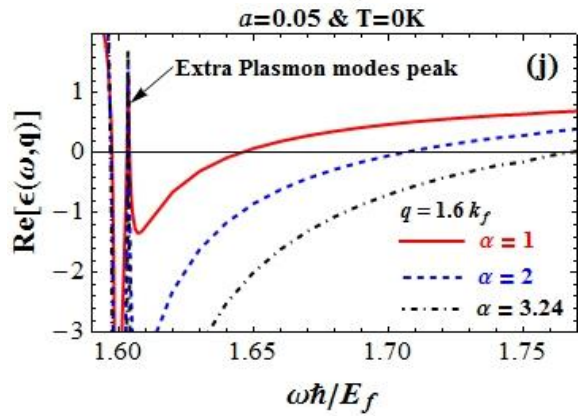

**(j)** $a=0.05$ & $T=0K$

**Figure 3** Density plot of single particle excitation (SPE) and plasmon dispersion (blue line) calculated at different gap and temperature values; (a) $a = 0.3$ & $T = 0.5T_f$, (b) $a = 0.3$ & $T = T_f$, (c) $a = 0.6$ & $T = 0.5T_f$, (d) $a = 0.6$ & $T = T_f$, (e) $a = 0.9$ & $T = 0.5T_f$ and (f) $a = 0.9$ & $T = T_f$. Green solid line shows zero-temperature plasmon dispersion. Panels (g) and (h) show the Plasmon dispersion vs normalised frequency calculated at $a = 0.05$ and $\delta = 10^{-4}$ for two different temperature values, and three different coupling constants, respectively. The real part of dielectric function vs frequency in panel (i) for different values of $q (= 1.5k_f, 1.6k_f$ & $1.7k_f)$ at $a = 0.05, \delta = 10^{-4}$, $T = 0K$ & $\alpha = 1$, and in panel (j) for different coupling constants $\alpha = (1, 2$ & $3.24)$ at $a = 0.05$, $\delta = 10^{-4}, T = 0K$ & $q = 1.6k_f$.

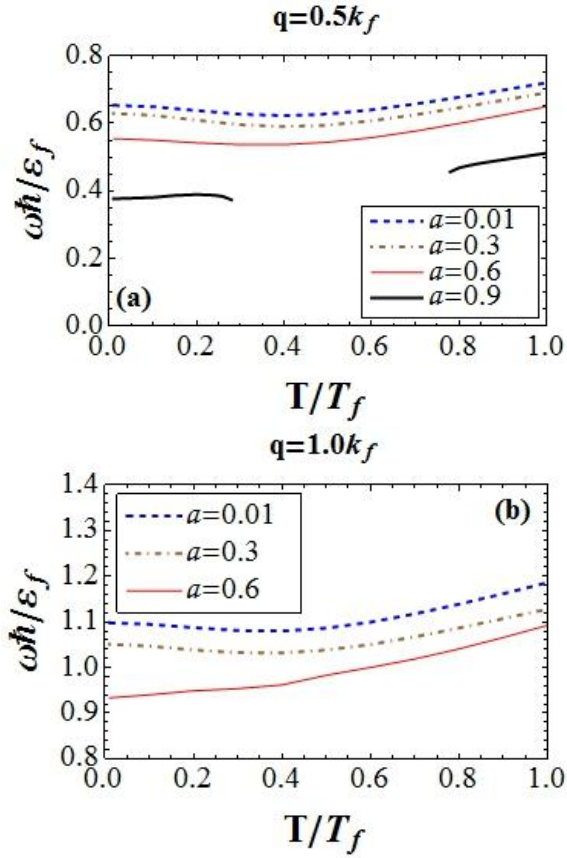

**Figure 4** Plasmon energy as a function of temperature for different gap values ($a = 0.01, 0.3, 0.6$ & $0.9$) at (a) $q = 0.5k_f$ and (b) $q = k_f$.

In Figs. 4(a)-4(b), the plasmon frequencies are plotted as a function of temperature for two different values of wave vector: $q = 0.5k_f$ and $q = k_f$. As can be noticed from Fig. 4(a), initially the plasmon frequency decreases with increasing temperature and attains minimum value at around $T = 0.5T_f$ and thereafter increases for higher values of $T$, for gap values of $a = 0.01, 0.3,$ & $0.6$, respectively. However, for $a = 0.9$, plasmon does not exist in the temperature range of $0.25T_f < T < 0.75T_f$ at

$q = 0.5\ k_f$. This disappearance is because the temperature suppressed plasmons from the begining enters the SPE intra band region in between two undamped plasmon region as also evident from the density plots of spectral weight function in Fig. 6(g). In the case of $q = k_f$ and $a = 0.9$, plasmons completely disappear as manifest from Fig. 4(b). This can be understood from Eq. (9) (or refers to Fig.5(b)), when the real part of dielectric function does not have zeros for any frequency at $q = k_f$.

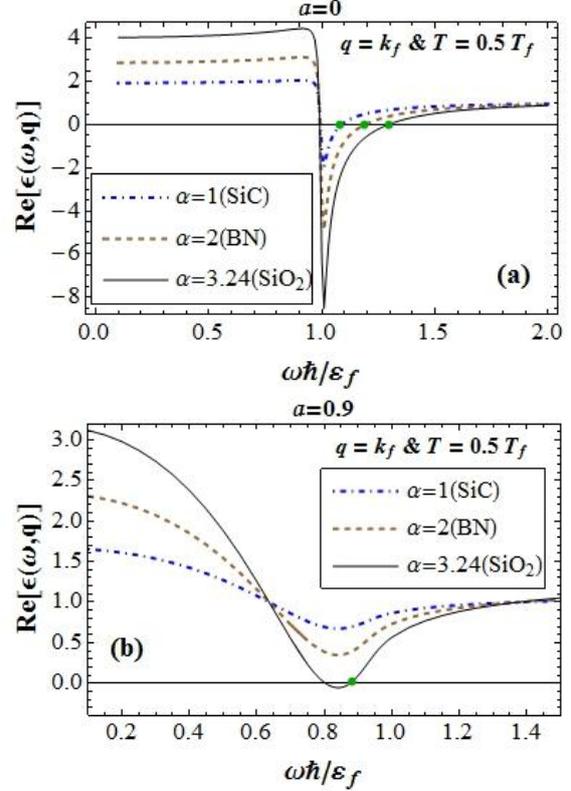

**Figure 5** Real part of the non-interacting dielectric function as a function of $\omega$ computed at $q = k_F$ and $T = 0.5T_F$ for different values of coupling constant ($\alpha = 1, 2$ & $3.24$) at (a) $a = 0$ and (b) $a = 0.9$. The green spot show the plasmon frequency $(\omega_p)$ at a given wave vector and temperature.

The computed real part of dielectric function as a function of $\omega$ at $q = k_F$ and $T = 0.5T_f$ for different values of coupling constant ($\alpha = 1, 2$ & $3.24$) at two values of gap ($a = 0$ & $0.9$) are depicted in Figs. 5(a)-5(b). As can be seen from Fig. 5(a), plasmon frequency increases with coupling constant at a given value of wave vector and temperature for gapless graphene. The Plasmon modes are observed at the following values of $\omega_p\hbar/\varepsilon_f \approx 1.0865, 1.1886$ & $1.2917$ for graphene sheet on SiC ($\alpha = 1$), BN ($\alpha = 2$) and SiO$_2$ ($\alpha = 3.24$) substrates, respectively. However in the case of gapped graphene, plasmon mode appears at $\omega_p\hbar/\varepsilon_f \approx 0.8764$ for graphene sheet on $SiO_2$ substrate ($\alpha = 3.24$) while the plasmon modes do not exist for lower subtrate values of BN ($\alpha = 2$) and SiC ($\alpha = 1$), as noticed from Fig. 5(b). This means that the substrate can be used a medium to tailor the plasmonic behaviour of graphene sheets.

**3.3 Electron energy loss function** Electron energy loss function is an important quantity which is directly measured in spectroscopic techniques like high resolution EELS. This quantity is proportional to the imaginary part of inverse dielectric function, $\text{Im}[1/\epsilon_{SLGG}(q,\omega,T)]$. In Figs. 6(a)-6(h) we show the density plots of the calculated SLGG electron energy loss function (i.e., $-\text{Im}[1/\epsilon_{SLGG}(q,\omega,T)]$) plotted in $(q,\omega)$ space and its comparison with the plasmon dispersion at zero-temperature (green solid line). In the density plots, the density scale shows the strength of the spectral mode. The loss function is directly proportional to the dynamical structure factor $S(q,\omega)$. The dynamical structure factor gives a direct measure of the spectral strength of the various elementary excitations [37]. The electronic excitations form the electron-hole continuum or single particle excitation (SPE) region in $(q,\omega)$ space. In Fig. 6, we also show both intraband and interband single-particle excitation (SPE) region and the surface plasmon region, which can be obtained from the non-zero value of the imaginary part of the total dielectric function, $\text{Im}[\epsilon_{SLGG}(q,\omega)] \neq 0$. When the plasmon curve enters the SPE continua at a given frequency and wave vector Landau damping takes place inside this region.

It can be noticed from Figs. 6(a)-6(b), for $a = 0.01$ and $T = 0.5T_f$ & $T = T_f$, respectively, the plasmon mode enters the interband SPE continuum where the Landau damping occurs and the the plasmons decay by creation of electron-hole pairs. Further, it can be noticed from Figs. 6(a)-6(b), for zero gap value the boundaries of intraband and interband SPE transitions merge. At $T = 0$ K and $a = 0.01$, we notice in Fig. 6(a) undamped plasmon mode up to $q \approx 0.95\,k_f$ and thereafter it enters the interband SPE region(green solid line). But with increasing gap values Figs. 6(c)–6(h) the undamped Plasmon region separates the boundaries of the two SPE regions with the separation widening with increasing gap values [25]. Moreover, from Figs. 6(e)-6(h) it can be observed from the plots that the SPE intra band region shrinks with the creation of undamped plasmon region at the higher end of $q$ values. At $a = 0.3$, plasmon mode shifts from the interband SPE region to the SPE gap and makes it undamped as shown in Fig. 6(c). Again plasmon mode shifts from the SPE gap to intraband SPE region for $q \gtrsim 1.3\,k_f$ ($a = 0.6$) and $q \gtrsim 0.7\,k_f$ ($a = 0.9$) as shown in Figs. 6(e) & 6(g) respectively.The plot in Fig. 6(g) for high band gap value $a = 0.9$ and $T = 0.5T_f$ corroborates the curve plotted in Fig. 4(a) for $a = 0.9$ at $q = 0.5\,k_f$ where the plasmon disappears in between $0.25T_f < T < 0.75T_f$ and after that reappears again. As pointed earlier, this plasmonic disappearance is because of the falling of the SPE intra band region in between the two undamped plasmonic regions.

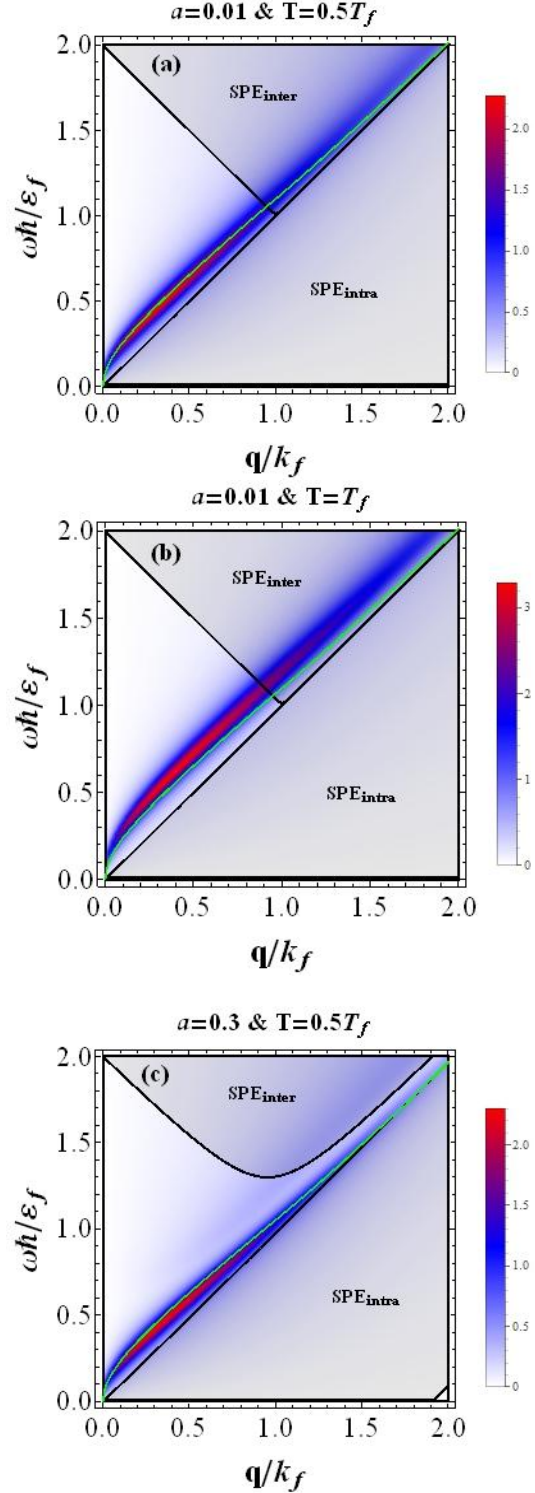

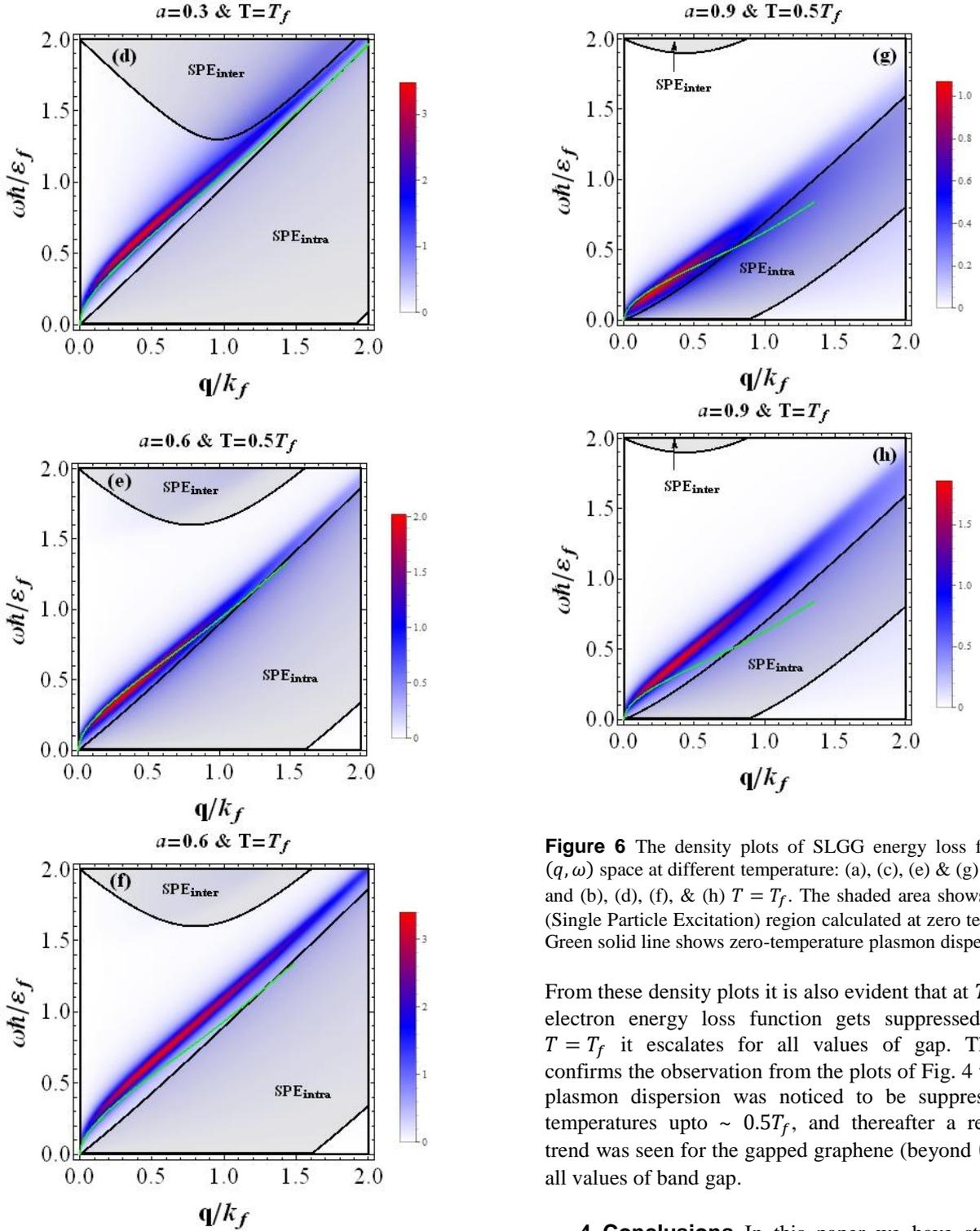

**Figure 6** The density plots of SLGG energy loss function in $(q, \omega)$ space at different temperature: (a), (c), (e) & (g) $T = 0.5T_f$ and (b), (d), (f), & (h) $T = T_f$. The shaded area shows the SPEs (Single Particle Excitation) region calculated at zero temperature. Green solid line shows zero-temperature plasmon dispersion.

From these density plots it is also evident that at $T = 0.5T_f$ electron energy loss function gets suppressed and for $T = T_f$ it escalates for all values of gap. This again confirms the observation from the plots of Fig. 4 where the plasmon dispersion was noticed to be suppressed with temperatures upto $\sim 0.5T_f$, and thereafter a reversal in trend was seen for the gapped graphene (beyond $0.5T_f$) for all values of band gap.

**4 Conclusions** In this paper we have studied the effects of temperatures and bandgap on the behaviour of non-interacting dynamical polarization (NDP) of doped Single layer gapped graphene within the RPA approximation. This NDP function has been then used to calculate the plasmon dispersions. We notice significant changes in plasmon dispersion curve due to temperature and bandgap. Our computed result shows an increase in plasmon dispersion with increasing wavevector similar to the case of gapless graphene. However, with increasing gap values the rate of increase of plasmon dispersion is seen to decline. We also find that plasmon dispersion decreases with temperature upto $\sim 0.5T_f$ similar to the gapless graphene case but a reversal in trend is seen for the gapped graphene beyond $0.5T_f$ for all values of bandgap. These observations are also confirmed by the density plots of electron energy loss function. The substrate also plays a prominent role in influencing the plasmonics behaviour. Overall the external and internal degrees of freedom endow gapped graphene with strange polarization behaviour and plasmons and can also serve as an effective means for its manipulation.

**Acknowledgements** Digish K. Patel acknowledges with thanks the UGC-BSR, New Delhi for financial support.